# Research On CODP Localization Decision Model Of Automotive Supply Chain Based On Delayed Manufacturing Strategy

Junchun Ding
*College of Engineering and Computer science*
*Syracuse University*
Syracuse, 13201, NY, United States
junchunding@gmail.com

*Abstract*—Under the market background of increasingly personalized product demand and compressed response cycle, the traditional manufacturing model with standardized mass production as the core has been difficult to meet the dual expectations of customers for differentiation and fast delivery. In order to improve the efficiency of resource allocation and market response, automobile manufacturers need to build a production system that takes into account cost and flexibility. Based on the delayed response manufacturing strategy, this study built an order response node configuration model suitable for automotive manufacturing scenarios, focusing on the positioning of order driven intervention points in the production process. The model comprehensively considers the structural cost changes brought by process adjustment, the dynamic characteristics of the changes of unit manufacturing cost and intermediate inventory cost at different stages with the location of nodes, and introduces delivery time constraints to embed time factors into the inventory decision logic to enhance the practicality of the model and the adaptation of realistic constraints. In terms of solution methods, this paper adopts function fitting and simulation analysis methods, combined with mathematical modeling tools, systematically describes the change trend of total cost, and verifies the rationality and effectiveness of the model structure and solution through actual enterprise cases. The research results provide a theoretical basis and decision support for automobile manufacturing enterprises to realize the synergy of flexible production and cost control in the environment of variable demand, and also provide an empirical reference for the implementation path and system optimization of subsequent relevant strategies.

## I. Introduction

With the automobile manufacturing industry facing increasingly fierce competition and increasingly complex market environment, the traditional large-scale production mode has gradually exposed its limitations, especially in the face of continuous compression of profit margins and rapid changes in market demand, the traditional production mode has been difficult to cope with consumers' strong desire for personalized and diversified needs. At the same time, consumers' requirements for delivery times are increasingly demanding, and the concern for cost performance is also increasing, which makes enterprises that cannot flexibly respond to market changes and timely meet customized needs, are easily replaced by other more adaptable competitors.

Only relying on standardized production mode has been unable to meet the needs of modern market and competitive pressure, we must seek more flexible and cost controllable production mode. In this context, the mass customization model came into being, which combines the high efficiency of mass production with the flexibility of personalized customization, and can not only meet the increasingly diversified consumer needs. At the same time, on the basis of effective cost control, it improves the competitiveness and flexibility of enterprises in the market. The effective implementation of this model depends on the rational use of delayed manufacturing strategy. By delaying the implementation time of customization, the delayed strategy helps enterprises to achieve accurate response to personalized demands while maintaining standardized production efficiency, thus maximizing inventory risk and optimizing production process.

The implementation of the delayed manufacturing strategy cannot be separated from the scientific determination of the customer order separation point (CODP), which marks the boundary between standardization and customization in the production process. A reasonable determination of the location of CODP can not only optimize the production cost structure of the enterprise, but also significantly shorten the delivery cycle and improve the overall operational efficiency. Therefore, the positioning of CODP plays a crucial role in production management. Choosing the right CODP location helps companies improve response speed and market adaptability while maintaining production flexibility.

By constructing a multi-dimensional CODP positioning decision model, this study provides a decision support tool with practical significance for enterprises, which can help them flexibly adjust production strategies in the face of rapid changes in the market, so as to improve the overall operation efficiency and market competitiveness. The decision model proposed in this study provides feasible technical path and optimization scheme for automobile manufacturing enterprises in the background of mass customization, and helps enterprises to stand out in the complex market environment.

## II. Relevant research

The key role of customer order decoupling points (CODP) in supply chain management has been extensively studied, especially in the context of customized production and delayed manufacturing strategies, whose optimization can effectively improve the efficiency and resilience of the supply chain.

X Hu [1] proposed a model based on multi-objective nonlinear optimization, and verified the effectiveness of the model in the scenario of multi-decoupling points through the improved NSGA-II algorithm, aiming to optimize the positioning of CODP by minimizing the cost of logistics service integrators and maximizing customer satisfaction. This study provides efficient decision support for customized logistics service supply chain. Based on this, X Hu [2] provides a multi-objective optimization model for CODP location optimization in the customized logistics service supply chain through an improved non-dominated sorting genetic algorithm (NSGA-II), successfully optimizing the cost of the logistics service integrator (LSI) and maximizing the overall satisfaction between the logistics service provider (LSP) and the customer.

M Harfeldt-Berg [3] deeply discussed the influence of different production modes on CODP selection through literature review, analyzed the operational differences between production to order (MTO) and production to inventory (MTS), and pointed out that different production modes directly affected the integration and customization ability of supply chain, thus affecting the allocation of CODP.

K Nguyen [4] studied the relationship between supply chain integration (SCI) strategy and CODP positioning in Vietnam's motorcycle supply chain, and found that demand volatility and market demand have a significant impact on the feasibility and performance of CODP allocation, emphasizing that SCI strategy and CODP allocation should be carried out simultaneously. To improve the responsiveness and flexibility of the supply chain. MS Muhammad [5] discussed the application of additive manufacturing (AM) in coping with the impact of COVID-19 on the automotive supply chain, and found that AM can effectively improve the resilience of the supply chain, especially in response to emergencies, which can speed up the production process and improve flexibility, and provide technical support for CODP optimization. H Carvalho [6] puts forward an indicator to evaluate the resilience of supply chain on time delivery, focuses on analyzing supply chain failure modes such as capacity shortage and material shortage, and emphasizes that enterprises can enhance the responsiveness of supply chain and maintain competitiveness by implementing resilience practices. S Zafar [7] introduces an automotive supply chain management framework based on blockchain technology, leveraging the Hyperledger Fabric platform to ensure data security and transparency. Blockchain technology provides information traceability for CODP management, ensuring the efficiency and security of supply chain management, especially in cross-company and cross-region supply chain collaboration.

TT Dang [8] combined spherical fuzzy AHP and grey COPRAS methods to propose a framework for assessing suppliers' adaptability to COVID-19. Research has shown that suppliers with emergency response mechanisms demonstrate greater resilience and sustainability in response to global crises, which is critical for the optimization of CODP. L Pinho Santos[9] explores the challenges of the automotive industry in driving sustainable supply chain transformation, proposing that through strategic alliances and innovative strategies, the automotive industry is able to respond to environmental pressures and drive sustainable development from "cradle to cradle", demonstrating the important role of sustainable supply chains in CODP optimization. W Sun [10] proposes a framework for CODP optimization through the study of supply chain resilience and enterprise responsiveness, emphasizes the coordination role of all links in the supply chain network, and proposes to enhance the adaptability of CODP through flexible supply chain and decentralized layout.

III. CONSTRUCTION OF CUSTOMER ORDER SEPARATION POINT POSITIONING MODEL BASED ON DELAY STRATEGY

*A Characteristics and delay strategy selection of mass customization in automobile manufacturing industry*

With the parallel evolution of the automotive industry towards high-quality development and personalized consumption, Oems have gradually shifted from the traditional capacity-driven model to the order-driven flexible manufacturing logic. Through outsourcing non-core manufacturing links to specialized supporting enterprises, dynamic reconstruction of production processes and optimization of resource allocation are realized. Oems focus on system integration, vehicle performance and architecture control. Parts manufacturing is coordinated by the supply chain to build an agile manufacturing ecosystem centered on user needs to cope with the rapidly changing market environment and diversified customization trends.

When CODP moves forward, it means that more processes are involved in the customized production stage, which leads to an extension of the customized production time, which also allows enterprises to better meet the requirements of customers for personalized needs. On the contrary, if CODP is moved back, more production processes will be classified as generalized production stages, which will reduce the degree to which customer customization needs are met. It can be seen that the specific location of CODP should be carefully selected, and the ideal location should be before the personalized feature process, which is usually located in the S process, and should not exceed this limit.

In the general production stage, as shown in Figure 1, enterprises usually adopt the centralized inventory strategy, and set the semi-finished products inventory buffer after the last general process of production to ensure the continuity of production. At the same time, in the customized production stage, because the product is processed according to the individual requirements of the specific customer, there is no need to set up the finished product inventory, but directly personalized distribution. In addition, the general production phase usually uses a push production model based on predictive information, while the customized production phase relies on a pull production driven by customer orders.

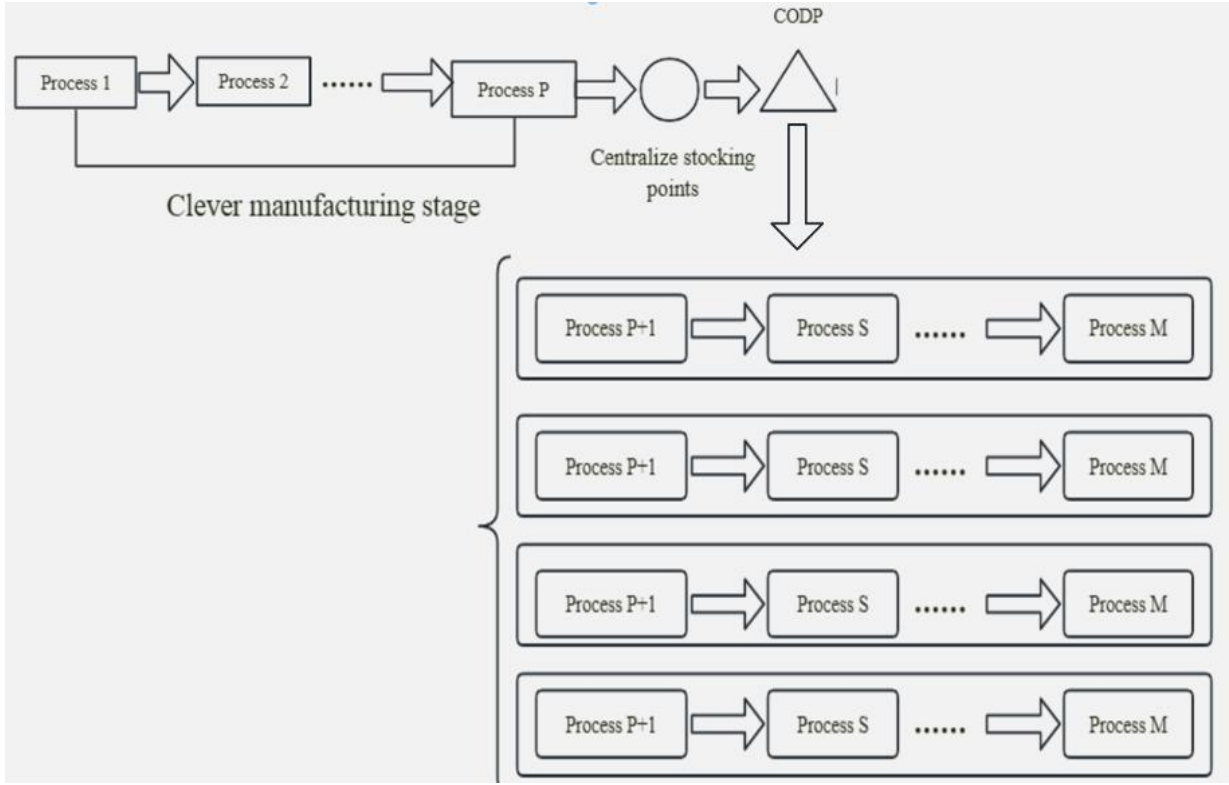

Figure 1. Production process model under mass customization with delay control

### B  Construction of customer order separation point positioning model

In the context of the continuous evolution of large-scale personalized manufacturing, vehicle enterprises need to scientifically set the startup node of customized operation in the process of implementing flexible production system, so as to meet the needs of diversified users and realize the effective constraint of manufacturing cost. Although the node setting too early helps to improve the customer response speed, it is easy to weaken the scale economy brought by the standardized process, resulting in the increase of inventory cost and resource waste. Too late may lead to frequent process structure adjustment, work-in-process accumulation and delivery delays, resulting in non-linear cost growth. In this paper, an optimization model is built around the impact of customized process nodes on manufacturing expenses, and the processing cost, inventory cost, process adjustment expenditure and resource allocation input are comprehensively considered. The objective function is defined in formula (1).

$$mine = \sum_{i=1}^{M}(C_{gi} + H_i + C_{ci} + R_i \cdot x_i) \quad (1)$$

Where $C_{gi}$ represents the standardized processing cost of modules, $H_i$ is the inventory cost of products in process, $C_{ci}$ is the incremental cost generated by personalized processing, $R_i \cdot x_i$ is the input for equipment adjustment and configuration, and variable $x_i$ is the decision quantity of whether to adjust the station. In order to meet the delivery time constraints and ensure that the customization task is completed in a reasonable stage, the model introduces constraints such as formula (2), manufacturing time constraints, and the customization operation stage constraints as formula (3). Constraints ensure that functional differentiation tasks can be completed while meeting timing and resource allocation, while balancing customization flexibility and production line load. Through quantitative modeling of flexible node locations, this study provides a systematic optimization path to guide enterprises to achieve synergy between cost minimization and response maximization in flexible transformation.

$$\sum_{k=p+1}^{M} T_{ik} \leq Di \quad \forall i \quad (2)$$

$$l < k < N \quad i < M \quad (3)$$

In order to more effectively meet the goals of personalized demand, delivery cycle and cost control, the research needs to introduce functional relationship on the basis of the original CODP positioning model to quantify the dependence between it and unit cost at each stage, and construct parameter mapping using actual production data, so as to realize quantitative analysis of the change trend of total manufacturing cost. Since the products belong to the same group, the unit cost in the general manufacturing stage can be regarded as a constant, while the average cost variable is set in the custom stage to reflect the differences of each manufacturing unit. In addition, considering the small difference in the processing time of the same process in the assembly line, the model can be uniformly treated as the average to simplify the analysis. In order to ensure the stability of the stock level, the replenishment cycle of the safe stock can be determined by calculating the average processing time of each process and expressed by the following formula (4).

$$L_s = \sum_{j=1}^{P} T_j \quad (4)$$

$L_s$ is the replenishment cycle of the safety stock, and $T_j$ is the processing time of the j process. Because the

processing time of each process may fluctuate. The uncertainty of inventory needs to be measured by statistical methods, which is embodied in formula (5). $\sigma_s$ is the standard deviation of process j processing time, and z is the coefficient set according to the level of service.

$$\sigma_s = z \cdot \sqrt{\sum_{j=1}^{P} \sigma_j^2} \qquad (5)$$

Given that different products share some common parts, inventory demand fluctuates over time. In order to accurately estimate inventory levels, enterprises need to use statistical models to quantify such fluctuations and further optimize inventory management strategies. By introducing multiple constraint factors such as production capacity, response time and service level, inventory allocation can be optimized, production efficiency can be improved, inventory costs can be effectively reduced, and enterprise operations can be optimized.

## IV. SOLVE THE CUSTOMER ORDER SEPARATION POINT POSITIONING MODEL BASED ON DELAY STRATEGY

### A Product delivery time constraint analysis

The positioning of CODP is carried out on the basis of meeting the delivery time and personalized needs, by analyzing the production process to identify the process involving personalized characteristics, and combined with the relationship between production time and delivery time to determine the scope of CODP. As CODP locations are adjusted, cost structures and inventory management strategies change. Since the production data are usually discrete, the unit cost function cannot be directly derived, so we use the curve fitting method to process it. By fitting the discrete cost data with SPSS software, the inventory of semi-finished products at buffer inventory points can be accurately calculated, so as to provide data support for CODP positioning. After further analyzing the relationship between unit cost and inventory, we construct a CODP positioning model based on delay strategy. Firstly, the limit of the number of processes is relaxed, and the cost function is optimized by MATLAB software to obtain the minimum cost value, and the best CODP positioning process is selected.

When the enterprise determines the CODP positioning process, to minimize the total manufacturing cost as the goal, it is necessary to comprehensively consider the delivery time constraints to ensure that the production cycle of customized products does not exceed the delivery time. When the delivery time is shorter than the shortest time required for customized processing, CODP cannot be located, and the assembly delay strategy is difficult to implement. When the delivery time is exactly equal to the minimum time, CODP shall be set in the first customization process; If the lead time is sufficient, the enterprise can flexibly select the positioning process in a wider range to optimize the allocation of resources. When the delivery time is between the customized processing time and the overall manufacturing time, it is necessary to determine a reasonable process interval through the function model, and achieve the collaborative optimization of cost and efficiency under the premise of meeting the delivery requirements.

### B Objective function analysis and optimal positioning process determination

When processing the trend extraction of discrete data, researchers usually adopt a continuous function model to approximate the distribution and change law of the data through reasonable parameter setting, which provides the basis for subsequent analysis. In order to improve the science of the model, analysts need to carry out systematic statistical diagnosis and feature extraction of the data before modeling, preliminarily determine the function form and construct multiple candidate models. Based on the distribution of data increments and the rate of change, the analyst may infer that the data fits different models such as linear, non-linear, or exponential.

During the implementation of the assembly delay strategy, the movement of CODP (customer demand demarcation point in customer-oriented flexible manufacturing system) has a complex impact on the cost of the overall manufacturing process. As CODP moves forward, the proportion of standardized production gradually increases, while the proportion of customized production correspondingly decreases. At this time, the complexity of the process and resource allocation involved in the manufacturing process is low, however, with the increase of the proportion of standardized production, the time and frequency of the production link have shown a certain increase, resulting in the rise of manufacturing costs. This change can be expressed by the following formula (6).

$$\frac{dC_2(p)}{dp} > 0 \qquad (6)$$

This indicates that in the case of CODP advance, the cost of the manufacturing stage will rise with the increase of production frequency and process complexity. On the other hand, the change of semi-finished goods inventory is closely related to the increase of total inventory cost. Under the centralized inventory strategy, the cost change of inventory area is represented by formula (7). $x_i^*$ represents standard inventory at each inventory stage, $h_i$ is unit inventory carrying cost, and $r_i$ is inventory turnover rate. According to the formula, the change of these variables indicates that the cost of inventory increases with the extension of the production cycle, so the change of inventory cost increases positively.

$$\Delta C_3(p) = \sum_{i=1}^{n} h_i \cdot (\frac{x_i^*}{2} + \alpha_i)/r_i \qquad (7)$$

When CODP is moved backwards, the proportion of product customization increases and the proportion of standardized production decreases, and while each individual process requirement in the manufacturing process may result in longer response times, the customized process is simplified and the diversity of processes is reduced, thus reducing the processing cost. In this case, the cost of customized production is gradually reduced as the production process is simplified.

### C Customer order separation point simulation and analysis

In the production process, the cost and production time of each process are analyzed according to the records of process transformation, manufacturing cost and production cycle. In these processes, some of the manufacturing costs are higher, especially in the early stages of manufacturing. In terms of production rhythm, some links need a longer time to be processed, especially some processes that require more delicate operations. In view of this phenomenon, the production efficiency can be further improved by adjusting the production process and optimizing the operation sequence, and the waste of ineffective time and resources can be reduced.

The data in Figure 2 and 3 show the cost and production time of each process. Through in-depth analysis, it is found that the key point of production optimization lies in the efficient integration of the early process and the reasonable connection of the later process. In view of these bottlenecks, the combination of customized production and general manufacturing can effectively improve the overall production efficiency, and reduce inventory occupation and semi-finished product backlog by optimizing inventory management, thereby reducing production costs and improving market response speed.

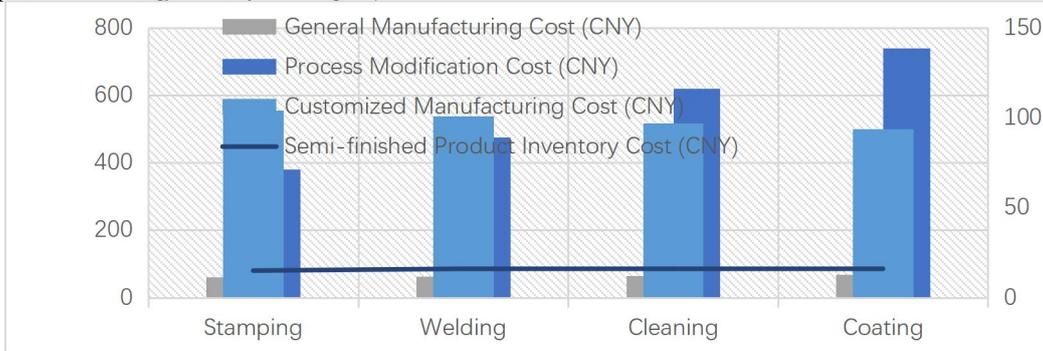

Figure 2. Generating cost statistics

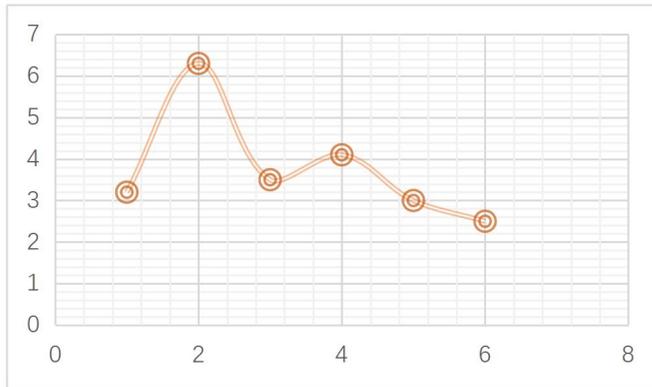

Figure 3. Production time data

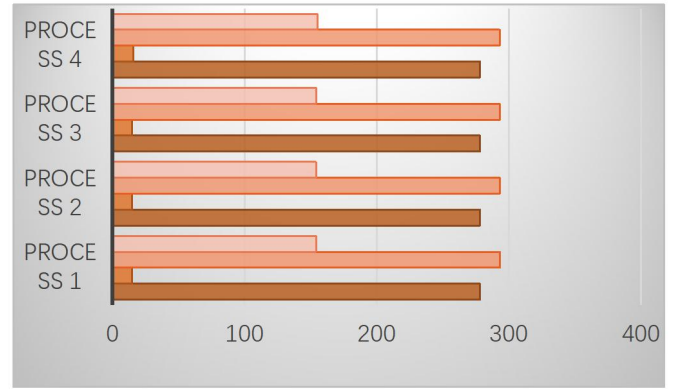

Figure 4. Analysis of buffer stock of semi-finished products at CODP with different positioning processes

By adopting centralized inventory management strategy and setting up semi-finished product inventory point to realize inventory control, using fixed inventory strategy to maintain the stability of inventory level. Stock checks are carried out regularly on a daily basis to ensure that the stock is always maintained at the predetermined level S and that 95% of the service level is achieved in this way. According to the calculated parameters, the safety stock level is determined to be 1.65, and combined with the time parameters in the production process, the safety stock and average stock of semi-finished products in different processes are further analyzed.

According to the data analysis results in Figure 4, when CODP is located at different process positions, the safe stock of semi-finished products does not change much in all processes, which is 15 and 16 units respectively, while the maximum stock of semi-finished products is always maintained at 293 units. Through these data, it is finally concluded that under the positioning of all processes, the average inventory of semi-finished products is stable at 154 units. In the analysis of process transformation and design costs, we found that with the gradual backward movement of CODP (customer order decoupling point), the transformation and design costs of each process showed an upward trend, and the specific data were shown in Figure 5. The data in the table reflect the cost changes of each process, and the index function model is used to fit, and the results show that the goodness of fit is 0.991, indicating that the model accurately describes the cost change law and has a high degree of fit.

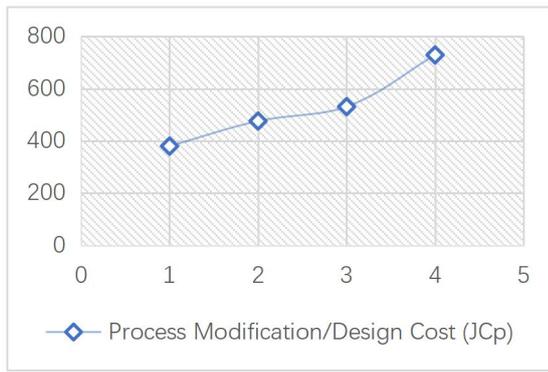

Figure 5. Statistical data of process modification/design cost

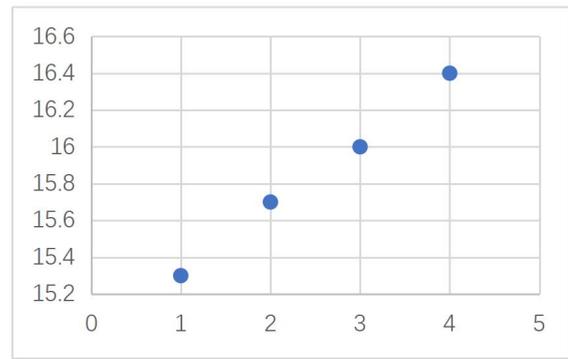

Figure 7. Data statistics of unit inventory cost of semi-finished products at buffer inventory points

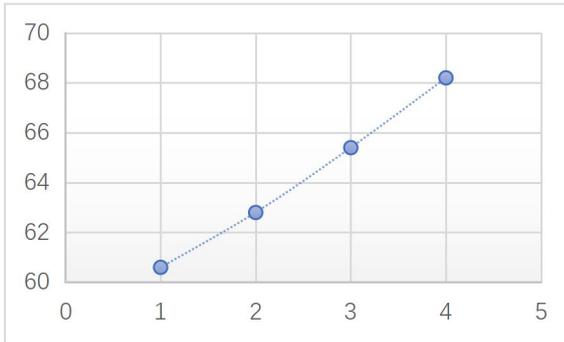

Figure 6. Statistical data of unit manufacturing cost in the general manufacturing stage

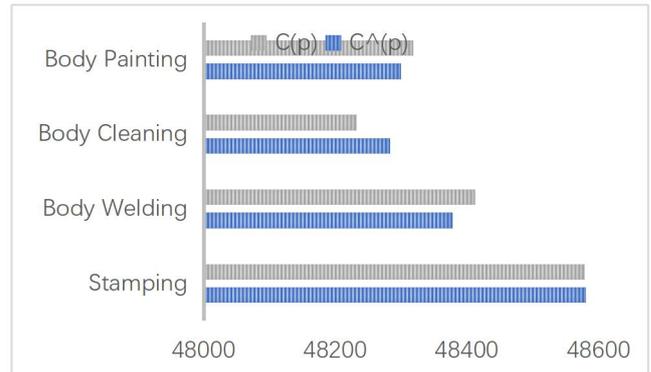

Figure 8. Total cost data statistics in the manufacturing stage

When analyzing unit production costs in the generic manufacturing phase, Figure 6 shows that production costs gradually increase as the CODP position moves back. The difference analysis results show that the second-order difference changes little and the trend is stable. The quadratic function model is selected to fit, and the results show that the goodness of fit reaches 100%. Moreover, the model significance test results show that the model can effectively describe the change trend of unit production cost. The comprehensive analysis shows that with the change of CODP position, process transformation and unit production cost show a consistent upward trend, and both exponential function and quadratic function models can effectively predict this trend.

In the customized manufacturing stage, with the gradual backward movement of the CODP positioning process, the unit production cost of each production process gradually decreases and shows a stable trend, as shown in Figure 7. In order to verify the reliability of the model, the goodness of fit and the significance of the equation are tested. The results show that the model can explain the change of the total cost in the manufacturing stage well and is statistically significant. Through these tests, the impact of each process on the overall cost is further confirmed, and the effectiveness of the model in practical application is verified. Figure 8 shows the cost data of different processes, which further proves the model's accurate prediction and reflection of the cost change trend.

Through the mathematical fitting and significance test of the manufacturing cost change model, it can be seen that different processes have different effects on the total cost. On this basis, the nonlinear function constructed expresses the trend of cost change with the process, and the key position corresponding to the minimum cost is determined by optimizing and solving. According to the further analysis of the impact of each process on the delivery cycle, it is found that the delay of the process can shorten the delivery time and improve the response efficiency. The enterprise should flexibly adjust the cutting node according to the requirements of the customer's delivery time. When the delivery time is very short, the whole process should be produced to ensure the delivery; when the delivery time is long, the optimal process can be cut in to minimize the cost. When the delivery time is in the middle, the second-best node can be selected to balance the cost and efficiency to realize the optimization of manufacturing decision.

## V. CONCLUSION AND PROSPECT

This study focuses on how to achieve reasonable scheduling of manufacturing resources and scientific allocation of process nodes under the background of increasing differentiation demand of automotive products. It proposes a key decision-making model that takes cost-effectiveness as the main line, takes into account production response speed and personalized satisfaction degree, and is simulated and verified by actual production data. It is proved that the model can effectively control the overall cost in the manufacturing stage under the premise of ensuring the delivery time and customization requirements. Future studies may further combine the upstream and downstream coordination mechanism to further explore the dynamic adjustment strategy under multi-process linkage conditions, and consider introducing real-time response mechanism in scenarios where multiple nodes coexist or demand changes frequently to improve system elasticity and adaptability, and explore how to build a flexible decision-making framework under the guidance of mixed strategies to coordinate the time difference between prediction and execution. It provides theoretical support and method path for improving the overall operation efficiency and customer responsiveness of complex manufacturing system.

REFERENCES

[1] Xiaojian H, Liangcheng X, Gang Y, et al. Multi-CODP decision models for supplier selection and order allocation in customized logistics service supply chain[J]. Neural computing & applications, 2024(19):36.

[2] Hu X, Xu L, Yao G, et al. Multi-CODP decision models for supplier selection and order allocation in customized logistics service supply chain[J]. Neural Computing and Applications, 2024, 36(19): 11097-11119.

[3] Harfeldt-Berg M. The role of the customer order decoupling point in operations and supply chain management[J]. 2024.

[4] Nguyen K, Nozomu K. Why and how firms conduct specific supply chain integration strategies? Considering the configurations of the customer order decoupling point and supply chain integration[D]. Tohoku University, 2023.

[5] Muhammad M S, Kerbache L, Elomri A. Potential of additive manufacturing for upstream automotive supply chains[C]//Supply chain forum: an international journal. Taylor & Francis, 2022, 23(1): 1-19.

[6] Carvalho H, Naghshineh B, Govindan K, et al. The resilience of on-time delivery to capacity and material shortages: An empirical investigation in the automotive supply chain[J]. Computers & Industrial Engineering, 2022, 171: 108375.

[7] Zafar S, Hassan S F U, Mohammad A S, et al. Implementation of a distributed framework for permissioned blockchain-based secure automotive supply chain management[J]. Sensors, 2022, 22(19): 7367.

[8] Dang T T, Nguyen N A T, Nguyen V T T, et al. A two-stage multi-criteria supplier selection model for sustainable automotive supply chain under uncertainty[J]. Axioms, 2022, 11(5): 228.

[9] Pinho Santos L, Proença J F. Developing return supply chain: A research on the automotive supply chain[J]. Sustainability, 2022, 14(11): 6587.

[10] Sun W, Li Y, Shi L. The Performance Evaluation and Resilience Analysis of Supply Chain Based on Logistics Network[J]. Conference papers, 2020, 000 (9181388). DOI: 10.23919 / CCC50068.2020.9189234.